# A New Family of High-Current Cyclotrons for Isotope Production

Daniel Winklehner[1], Jose R. Alonso[1*], Janet Conrad[1]

[1]*Massachusetts Institute of Technology: 77 Massachusetts Avenue, Cambridge, MA, 02139, USA*

## Abstract

We are developing a high-current cyclotron as a driver for the IsoDAR neutrino experiment. It accelerates 5 mA $H_2^+$ to 60 MeV/amu, after which the electron is removed to produce a 10 mA, 60 MeV proton beam. The enabling innovations that offset space-charge effects occur at injection and in the first few turns, allowing one to construct cyclotrons with energies ranging from below 5 MeV up to 60 MeV/amu, or possibly higher, with the same performance for accelerated ions with Q/A = 0.5 ($H_2^+$, $D^+$, $He^{++}$, ...). In this paper, we discuss the possible uses of such cyclotrons for isotope production, including production of long-lived generator parents ($^{68}$Ga, $^{44}$Ti, $^{82}$Sr,...), as well as intense fast neutron beams from deuteron breakup for (n,2n) production of isotopes like $^{225}$Ac.

## Keywords

## Introduction

The successful production of radioisotopes for medical use involves a symbiotic relationship between several communities each with very significant challenges that must be met. Isotope production involving, in our case, accelerators, pushes cyclotrons or other accelerating technology to the highest currents, and lowest beam loss to mitigate activation and allow maintenance. The high powers of the cyclotron beams, striving for the highest production rates, pushes target technology and the ability to dissipate the very significant heat deposited in the targets. Once irradiated, the targets must be processed to extract the produced activity in a safe and efficient way in the presence of extreme radioactivity. Meeting the challenges in each area pushes the boundaries of knowledge and technology in their respective fields and requires excellent communication and cooperation of the diverse teams. In all, the progress and impact of radioisotopes on medicine and human health have been outstanding.

This paper will focus primarily on the accelerator part of the equation, and will touch on some of the targetry issues, but the extraction and purification areas are outside the scope of this study. Extensive literature is available for those interested in this area.

Here we introduce the isotope community to a new family of cyclotrons developed as part of the particle-physics neutrino program. By virtue of their significantly higher beam currents, these novel accelerators have the potential to produce large quantities of valuable and much needed isotopes in the future. Some R&D is required to allow application to isotope production, and we highlight the primary remaining questions that must be addressed. We argue that the step-forward for the industry in using these machines is well worth the investment needed to address the outstanding issues.

To make this case, we begin with an overview of the importance of cyclotrons to isotope production now, and the value of increasing cyclotron beam intensities in the future. We then consider the specific design that was developed for neutrino physics that can be applied to high intensity isotope production. We hope this information will stimulate research programs that allow for cyclotrons with this new, high-intensity design to make a positive impact on the isotope community.

## Cyclotrons Are Key for Isotope Production

Cyclotrons and reactors are complementary instruments for producing medically relevant radioisotopes. Reactors are favored for neutron-rich products while reactions with cyclotron beams favor the proton-rich side of the nuclide chart [1, 2, 3, 4]. In particular, many of these proton-rich nuclei decay by positron ($\beta^+$) emission and are useful PET isotopes. A major advantage of cyclotrons over reactors is the ability to more easily produce carrier-free reaction products. The $(n,\gamma)$ reactions typical of reactor irradiations result in a product almost always of the same atomic species as the target. As a result, separation of the product from the target material is difficult, and placing the produced isotope onto a carrier molecule results in very low specific activity for the resulting tagged pharmaceutical. Moreover, cyclotrons permit flexibility in beam energies that enables tailoring a cyclotron to specific reactions. Low energy cyclotrons (10-15 MeV) are best for $(p,n)$ reactions, being small and compact enough to be emplaced in hospital settings, close to the end-use point. Medium-energy machines, in the 30-40 MeV range, as well as higher-energy machines, above 60 MeV, offer a wider range of reaction possibilities, and are often used in regional production centers, from which processing and distribution networks enable delivery of the desired isotopes to end-use points.

End-use isotopes will mostly have short half-lives, especially those used in diagnostic procedures. These diagnostic procedures typically require no more than an hour to collect the required data, and any remaining isotope decay after this time contributes unnecessary radiation dosage to the patient. Short half-lives create a distribution problem unless the isotope can be produced at the end-use site, perhaps the best example of this is $^{18}$F ($T_{1/2} = $ 110 minutes), produced mostly by a $(p, n)$ reaction on $^{18}$O with compact 10-15 MeV cyclotrons. A method of mitigating this distribution difficulty is by using "generators" where a long-lived parent, produced and shipped to the end-use point, decays into a short-lived daughter that can be "milked" (typically by elution) from the parent and used in the patient [1]. The best-known example of this is the $^{99}$Mo/$^{99m}$Tc generator, the $^{99}$Mo parent has a 66-hour half-life, is a prominent fission fragment from $^{235}$U, and the $^{99m}$Tc half-life is 6 hours, suitable for SPECT imaging [5]. Another very useful generator pair is $^{68}$Ge/$^{68}$Ga [6]. $^{68}$Ga is a 68-minute positron emitter (89% $\beta^+$, 11% electron capture), that is being widely used when available, for PET studies. The parent $^{68}$Ge has a 271-day half-life and is produced by proton bombardments on gallium. Gallium has two stable isotopes, $^{69}$Ga (60%) and $^{71}$Ga (40%). A cyclotron-produced proton beam of 60 MeV on a thick target of natural gallium can produce the $^{68}$Ge parent isotope either through the (p,2n) channel on $^{69}$Ga, or a (p,4n) reaction on $^{71}$Ga [7,8].

## Cyclotrons for isotope production today

The uses described above have generated a thriving accelerator industry. Cyclotrons used for isotope production are marketed in North America primarily by two vendors: IBA (Belgium) [9] and ACSI (Canada) [10]. GE also offers a PET cyclotron system but has no higher-energy machines [11]. ACSI cyclotrons are all below 30 MeV, while IBA, in addition to covering this energy range, also offers the C-70, producing proton beams of 70 MeV. To make proton beams, all commercial cyclotrons from IBA and ACSI accelerate H$^-$ ions, extracting the beam with a stripping foil whose position can be adjusted inside the cyclotron to provide a range of extracted proton energies. The maximum current from these existing machines is of the order of 1 mA due to limitations from beam capture and extraction [8, 12]. The other major commercial vendor is Sumitomo [13], with extensive experience in all manner of cyclotrons, from the collection of heavy-ion machines at RIKEN [14], to a high-current (1 mA) cyclotron for BNCT [15].

Specialized research machines at historical accelerator laboratories offer higher energy beams for isotope production, e.g. (in North America) TRIUMF [16], Texas A&M [17], Brookhaven National Lab, [18] and Los Alamos National Lab [19]. However, in the US, the Department of Energy limits these programs to produce only those isotopes that do not compete with on-market facilities. These isotope production programs are also generally low priority compared to the physics programs at the laboratories, limiting run-time. For example, the Los Alamos facility had no isotope beam running in 2023. As a result, we concentrate in this paper on how to increase isotope production through commercial facilities.

## Limits to existing commercial cyclotron current

H⁻ cyclotrons usually inject beam with an energy of 15 keV, and because of poor capture efficiency must provide adequate current to overcome losses in the central region. For 1 mA of extracted beam, the typical injection current is around 10 mA. This current is steady-state, and because of the very high space-charge forces, bunching to provide better beam utilization by increasing the particle density within the narrow phase acceptance of the cyclotron RF system is essentially useless [20]. The high beam losses in the central region also lead to serious erosion damage, requiring rebuilding of the central region every few years.

The other major limitation for H⁻ cyclotrons is the extreme heating of the extraction foils. At the extraction radius the H⁻ ions impinge on the edge of a thin carbon foil, which strips the two electrons from the H⁻ ion. The now-bare proton is bent in the opposite direction in the cyclotron magnetic field, and cleanly leaves the cyclotron. However, the two "convoy" electrons are tightly bent in this field, and pass repeatedly through the stripping foil until all their energy is lost. In a 30 MeV 1 mA cyclotron the electrons in a 0.5 mA beam (two extraction foils are usually used, splitting the beam power equally between each) deposit approximately 17 watts of power, raising the temperature to almost 2000 K. Experimental data indicate that at higher temperatures the foil lifetime is severely affected. Thus, higher currents can only be achieved by preventing the convoy electrons from reaching the foil, placing the foil in zero-field regions or introducing gradients to spiral the electrons away from the foil. Both are difficult to achieve inside the cyclotron.

## The road to increased capacity

Is there value in increasing the current-delivery capacity for isotope-producing cyclotrons beyond 1 mA, that might be achieved by reimagining the cyclotron? We believe the answer is yes, especially for the practical production of long-lived generator parents. The activity produced per microamp-hour of irradiation time is inversely related to the lifetime of the product. For long-lived generator parents such as 25.5-day ⁸²Sr, 271-day ⁶⁸Ge, or -- even more so -- 59-year ⁴⁴Ti, making a generator of adequate strength must be accomplished either by extended running or increased beam-on-target. An increase in beam current by a factor of 10 or more decreases the irradiation time by this same factor.

While normal market forces might not drive cyclotron builders to develop these higher intensities, needs from particle-physics research -- in our case driven by the requirement for the highest-possible neutrino fluxes for the IsoDAR experiment [21, 22] -- have provided the impetus for our innovations. Our development, described below, is a cyclotron that accelerates $H_2^+$ to 60 MeV/amu, where these ions are extracted, and the electron is stripped from the molecule in the transport line. The result is a 60-MeV 10 mA proton beam, a driver for a very intense source of neutrinos. We describe the IsoDAR experiment in a subsequent section.

We believe that these higher-current beams have excellent potential applicability to isotope production. But at the same time, we realize that the high-power beams raise major targetry issues. The power-handling capability of existing targets is, at best, matched to the 1 mA currents of present-day cyclotrons. Indeed, most present-day targets cannot absorb that level of current because efficient cooling is difficult to provide. However, we feel strongly that this is a practical engineering limit not a hard-physical one. While not feasible at the present time, future developments, either in power-handling of targets, or in the ability to split the beam between several target stations, could increase the utilization of the available beam from a 10-mA cyclotron. A concept for effectively splitting an extracted molecular $H_2^+$ beam has been discussed previously [12], by inserting a stripping foil at the edge of the external $H_2^+$ beam, generating protons, that are more sharply bent in a dipole magnet just behind the stripper. Beam on target is controlled by the amount the foil is inserted into the beam. The $H_2^+$ beam can be further utilized by repeating this stripping maneuver in downstream stations. Several beamlets of the order of 0.5 to 1 mA could thus be produced, suitable for present-day targets, while targets capable of absorbing more beam power are being developed. This is one example of how the development of high-current machines could be an opportunity to drive development of innovative solutions to the targetry problem.

### Beams other than protons

Although the IsoDAR cyclotron was developed to accelerate an $H_2^+$ primary beam, any ion with a charge-to-mass ratio of 1 to 2 can be equally well accelerated with relatively modest changes to the tuning. So, in addition to protons (as $H_2^+$ ions), beams of $D^+$, $He^{++}$, even $C^{6+}$ ions can be produced opening more opportunities for isotope production. There are slight nuclear mass differences between the various ions, as shown in Table 1, but magnetic field profiles can be easily adjusted for these small differences to preserve isochronicity for all these ions. Isochronicity is required for high-current cyclotrons to enable all the orbits to be populated with beam bunches throughout the entire acceleration process.

**Table 1** Deviations in q/A values from nuclear mass differences

| Ion | Charge | Mass (MeV) | q/A Normalized to $H_2^+$ = 0.5 | % Difference |
|---|---|---|---|---|
| $H_2^+$ | 1 | 1877.051 | 0.5 | --- |
| Deuteron | 1 | 1875.61 | 0.500384 | 0.08% |
| Alpha | 2 | 3727.38 | 0.503594 | 0.72% |

Deuterons and alpha particles have long been used for isotope production [23, 24], either directly or, in the case of deuterons, by converting the beam into a fast neutron beam by stripping the neutron from the deuteron nucleus [25, 26, 27]. Fully stripped carbon is a novel ion, now commercially available from the most powerful ECR sources manufactured by Pantechnik [28]. There is an interesting possibility of directly producing alpha-emitting $^{149}$Tb with a $^{141}$Pr target (this element has only one stable isotope), via the $^{141}$Pr($^{12}$C,4n)$^{149}$Tb reaction. The total energy needed is around 70 MeV (5.8 MeV/amu), and the peak of the 4n cross section should be in the range of 100 mb. There are only two problems: a) the current of carbon 6+ from the ion source is only a few hundred particle nanoamps, and b) dE/dx for heavy ions has a $Z^2$/A dependence, dropping the effective thickness of a target to cover the excitation function by about a factor of 3, so it will contain substantially less material. Nonetheless, with these parameters a yield of about 40 kBq/hr might be obtained, perhaps sufficient for studying the dynamics of the reaction, while waiting for ion sources to be developed with higher currents.

The He$^{++}$ current from the Pantechnik source is 1 mA (500 particle microamps), we anticipate that current on target would be about 250 particle microamps. This would be substantially above the 30 microamps currently used at RIKEN in their dedicated [211]At production cyclotron [29].

Modifications of the injection systems that will be described below could enable these cyclotrons to utilize lower charge states from the ion source, resulting in substantially higher currents for heavier ions. Such possibilities are beyond the scope of this paper and will be addressed in separate studies.

But perhaps the most interesting possibility is the use of fast neutrons from the dissociation of deuterons with top energies of the order of 40 to 50 MeV, with neutron energies covering the range from 25 MeV downwards.

The use of fast neutrons from the breakup of accelerated deuterons is well-known and has been extensively studied [25, 26, 27]. We would like to focus in this paper on the production of [225]Ac via the (n,2n) reaction.

## [225]Ac production as a key example

One of today's most promising therapeutic isotopes, [225]Ac with its decay chain including four alpha particles, has already been shown to be highly effective in the treatment of castration-resistant metastatic prostate cancer [30]. Development of other treatments is ongoing, but the pace of these developments is hampered by the scarcity of the isotope, and the difficulty in its production.

The most accessible method using 20-30 MeV cyclotrons is the (p,2n) reaction on [226]Ra producing [225]Ac directly [31]. This requires thin metallic radium targets that must be dissolved and reconstituted after every irradiation, as well as sophisticated hot-chemistry facilities to work with the 1600-year $T_{1/2}$ [226]Ra target material. A target covering the peak of the (p,2n) excitation function would have an activity level of around 20 gigabecquerel (0.5 curies}.

The US Department of Energy has developed a process using high-energy protons bombarding natural thorium to produce [225]Ac from spallation reactions [32]. The cross section is low, but the high currents from the accelerators at Los Alamos and Brookhaven enable a reasonably good yield of [225]Ac. TRIUMF is also using its 500 MeV cyclotron beam to produce [225]Ac via this spallation reaction [33]. The principal drawback of this technique is the production of [227]Ac whose long half-life renders it an unacceptable contaminant. While Robertson at TRIUMF has developed a chemical process for isolating [225]Ac based on the short half-life of [227]Ra [34], this process comes at the cost of somewhat reduced yield of the desired [225]Ac. Lack of continuous year-round access to the high-energy accelerators also limits the usefulness of this channel.

The (γ,n) channel can be effectively utilized to produce [225]Ra from [226]Ra. The 15-day [225]Ra β- decays into [225]Ac, and so is a good generator for the desired Ac isotope and provides a degree of independence for a continuous supply of [225]Ac somewhat decoupled from accelerator schedules. As an example of a commercial application, NorthStar Medical Isotopes [35] is utilizing 10 MeV IBA Rhodotrons [36], high-power electron accelerators and bremsstrahlung converters to produce the gamma fields to irradiate the [226]Ra targets. To increase yield, they employ two Rhodotrons with gamma fields irradiating the target from both sides. The (γ,n) cross section in the 10-20 MeV range is in the few hundred millibarn range, but the production yield is limited by the concentration of the gamma flux. This process has the potential for improving the world's supply of [225]Ac free from [227]Ac contamination.

Deuterons can potentially provide substantially higher quantities of [225]Ac. The (n,2n) reaction cross section on radium in the energy range between 10 and 20 MeV has a very large cross section, shown in Fig. 1 to have a peak at 2.3 barns [37]. Covering the full excitation function, from neutron energies of about 17 MeV down to 7 MeV still provides an average cross section above 1 barn. One can view the neutron as entering the nucleus and elastically scattering off a bound neutron, with this second neutron recoiling out of the nucleus. This is a very efficient direct-reaction process, and in fact the cross section is a substantial fraction of its geometric value.

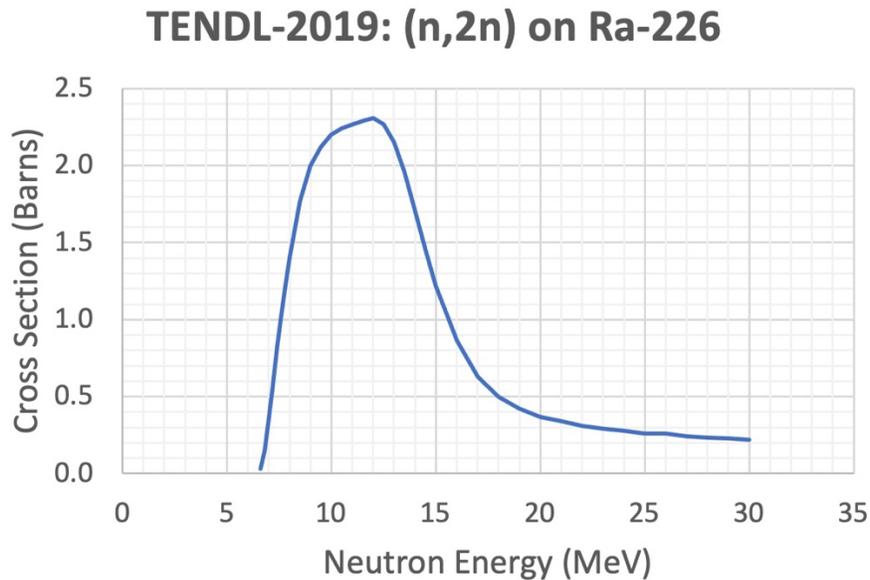

**Fig. 1** TENDL-2019 cross section for (n,2n) on [226]Ra [37]. These calculations have best agreement with experimental data [27]

In a similar vein, the cross section is also quite high for the breakup of a deuteron yielding a forward-going fast neutron with approximately the velocity of the deuteron at the moment of its interaction with a beryllium nucleus. Ref. [27] presents a calculation showing that for deuterons of 40 MeV entering a thick beryllium target, approximately 9% are converted into forward-going neutrons. The results of this experiment are consistent with a very low fraction of slow neutrons in the irradiation field which could lead to (n,γ) reactions in the radium target. In fact the author reports being unable to detect any [227]Ac contamination even after steps to improve the sensitivity for detecting this isotope. Nevertheless, this conclusion needs to be tested with higher intensity deuterium beams, where sensitivity to [227]Ac contamination might be greater.

Production of [225]Ra (the precursor to [225]Ac) would occur by placing the radium target downstream of the Be breakup target, reasonably close (a few cm) behind the Be. This is close enough to have the highest density of fast neutrons, but not so close as to interfere with the cooling of the Be target. This concept is illustrated in Fig. 2. Note, the Ra does not need to be in metallic form, nor does it need to be thin, so a chemical form can be chosen from which the Ac daughter can be suitably extracted. Based on estimates from Ref. [27] for scaling the experimental setup developed at the LBNL 88-Inch cyclotron to a realistic production facility with an optimized geometry for efficient utilization of the fast neutrons, we predict that our cyclotron with a 5 mA of deuteron beam could produce above 0.4 terabecquerel (10 curies) of [225]Ac per day of running.

The experimental measurements detailed in Ref. [27] , although considerably scaled back in deuteron beam intensity due to radiation protection and shielding limits at the LBNL 88-inch cyclotron, demonstrate the feasibility and practicality of this mechanism for producing [225]Ac. Scaling up and optimizing targets, both for deuteron breakup and radium irradiation, hold significant promise for greatly expanding the availability of this isotope for clinical applications, but experimental verification of this must be demonstrated.

In summary, high-power deuteron cyclotrons producing intense secondary neutron beams can be economical vehicles for generating valuable isotopes. The targets that convert the deuteron beam into a field of fast neutrons are totally different from those used for actual production of the medical isotopes and can be built more easily to withstand the extreme beam powers available. Furthermore, as the secondary neutron beams carry no charge, their energy loss in the isotope-producing media is low so constraints on the amount of material used are low and there is more flexibility in its chemical form, and thus allow for optimized production with little consideration for target thickness or the heating of this material.

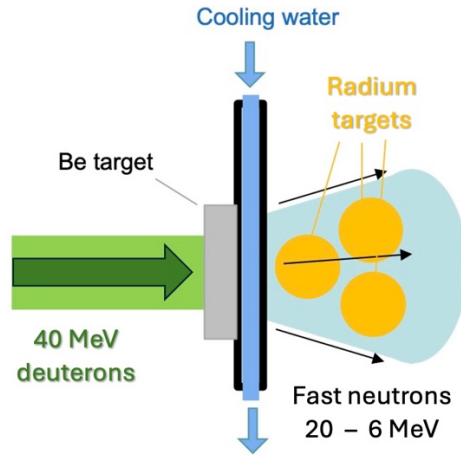

**Fig. 2** Schematic of experimental setup for fast neutrons behind beryllium target. Incident ~40 MeV deuterons break up with forward going neutron having velocity of deuteron at instant of breakup. Target thickness is such that breakup event at the back side of target produces ~6 MeV neutrons. This covers the whole (n,2n) excitation function. The residual beam particles stop in the cooling water. Radium targets are placed in fast neutron field.

### *Neutron-production target considerations*

[225]Ac production by new high-powered cyclotrons raises one example of the types of R&D questions that must be explored for the industry to make use of a new generation of machines. However, one important issue -- the target used for the deuterium breakup -- can build on developments that have already been well documented. Many different materials can be used, depending on the specific characteristics of the desired neutron field.

High power beryllium targets are in widespread use in this beam-energy range for BNCT neutron production [38, 39]. There are two considerations important to the longevity and survival of these targets: a) The power density on the target should be kept at or below 2 kW/cm$^2$, and b) the stopping point of the protons must not be in beryllium to prevent blistering of the beryllium due to accumulation of stopped protons as hydrogen gas which is insoluble in beryllium [40]. As will be described below, for the IsoDAR experiment the former is accomplished by spreading the beam out over a large area, in our case 20 cm diameter, accomplished by wobbling the beam with sinusoidally-driven magnets in the beamline upstream of the target. The latter is in principle straightforward to do in a flat target where one makes the target sufficiently thin, so the protons penetrate through and stop in the cooling water behind the target, or in the IsoDAR case by carefully modulating the wobbling pattern of the beam to keep protons from stopping in the hemispherical beryllium target shells.

While beryllium is a satisfactory material, and its use has been optimized for the IsoDAR experiment, other materials have been explored for producing intense fast-neutron fields, including liquid lithium [41], and the substantially more refractory carbon [26].

## The IsoDAR experiment and its cyclotron

The driving force behind our work to design this cyclotron that we are now introducing to the isotope community has been the IsoDAR (Isotope Decay At Rest) neutrino experiment [21, 22, 42].  Neutrinos have proven enigmatic, and incredibly interesting to the particle-physics community [43] and the subject multiple Nobel prizes (Lederman, Schwartz, Steinberger, 1988; Reines, 1995; Davis, Koshiba, 2002; McDonald, Kajita, 2015)  The observation that neutrinos oscillate between the three known "Standard Model" flavors [44, 45] that is not predicted by the Standard Model points to the neutrino as being a very likely candidate for the discovery of "New Physics," or "BSM" (Beyond the Standard Model) physics [46], one of the Holy Grails in particle physics for about 50 years now. Because the interaction cross sections of neutrinos are so incredibly low (in the range of $10^{-42}$ cm$^2$), experiments require large detectors (with sensitive volumes measured in kilotons), ideally in very quiet environments (e.g., deep underground).

### *A very short introduction to IsoDAR*

To understand our design choices for the cyclotrons under discussion, it may be useful to place the context for which the 60 MeV/amu $H_2^+$ machine was designed.  The IsoDAR experiment will place an intense source of antineutrinos a few meters from a new 2.3 kiloton liquid scintillator detector to be built in an existing cavern in the Yemilab underground laboratory one kilometer below the surface in the Eastern mountains of South

Korea [47]. To produce this intense flux of neutrinos, the IsoDAR cyclotron will deliver 10 mA of 60 MeV protons (600 kW) to a beryllium and heavy-water target, producing almost a mole of neutrons per year. These neutrons stream into a near-spherical sleeve surrounding the target containing a mixture of beryllium (75%) and highly enriched (>99.99%) $^7$Li. The lithium captures the neutrons producing $^8$Li, whose beta decay (with an 840-millisecond half-life) produces an electron antineutrino with very high endpoint energy, ideally suited for definitive experiments relating to neutrinos. Recent studies [22] have indicated this beam-target combination can also be useful in exploring a wide range of new physics. Because the source of neutrinos is radioactive decay (of $^8$Li), it could be considered that it is really a radioactive-source experiment, with a 1.5 petabecquerel (40 kilocurie) source that is constantly replenished by the accelerator beam.

The experiment is shown schematically in Fig. 3, Yemilab (3b) accesses the underground infrastructure of the Handuk iron mine, the largest in South Korea, while 3a shows the layout of the IsoDAR experiment. The cyclotron, in a specially constructed cavern, produces the $H_2^+$ beam that is stripped and transported as protons to the target close to the planned 2.3 kiloton liquid scintillator detector. Fig. 4 is a rendition of the target area showing the wobbled beam impinging on nested beryllium hemispheres cooled by heavy water. The target is surrounded by the 1.6 m diameter shell containing the $^9$Be-$^7$Li mixture. The beam is bent through 180° so it impinges the target going away from the detector, to minimize the flux of fast neutrons directed towards the detector. 7 meters of steel between the target and the detector keep the flux of fast neutrons (> 3MeV) to less than natural background at this depth, a few per year in the entire fiducial volume.

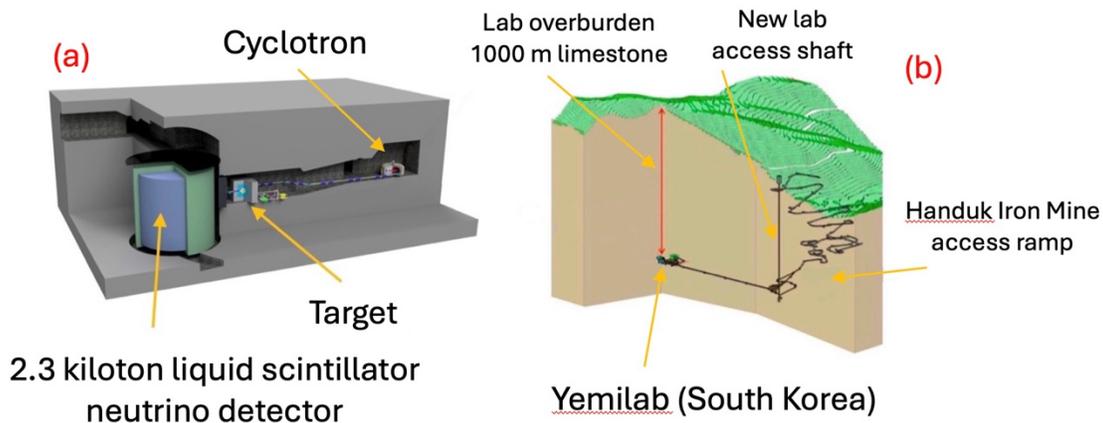

**Fig. 3** The IsoDAR experiment will be deployed in existing caverns in the newly completed Yemilab underground laboratory adjacent to the Handuk iron mine in Eastern South Korea. (3b adapted from Seo [47]).

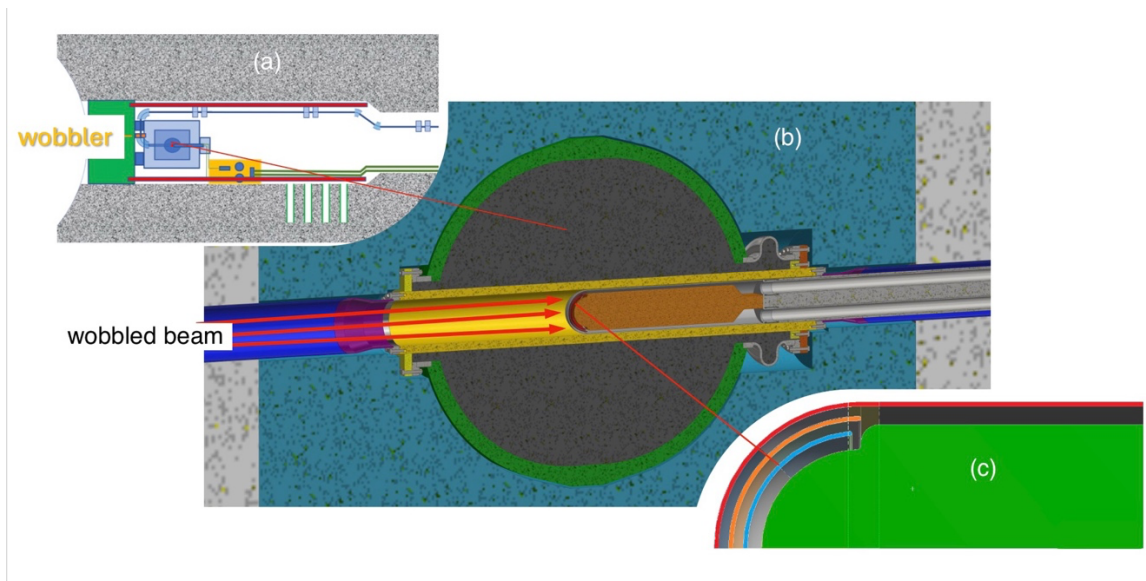

**Fig. 4** IsoDAR target area and details. (a) Layout of target cavern. The beam wobbler is the orange element in the transverse part of the beam line. The large green block is part of the 7 meters of steel shielding to attenuate any remaining fast neutrons. (b) Target and sleeve. The target is at the end of a 3 meter-long "torpedo" inserted into the beamline vacuum tube from the downstream end, configured for easy exchange and remote handling. The target is at the center of the ~800 mm radius beryllium shell containing the 75% Be and 25% highly enriched $^7$Li mixture. (c) Detail of target: Nested beryllium hemispheres, each 3 mm thick, cooled by flowing $D_2O$. The diameter of the torpedo is 200 mm.

In the following sections we detail the innovations that lead to the current increases in our cyclotron.

## IsoDAR innovation 1: alternative thinking on the ion source

Multiple innovations have been brought to the IsoDAR design, starting with reconsideration of the best choice of ion source. Space charge must be managed to reach higher currents, of which the principal area of concern is in the transport from the ion source and injection into the central region, where the ion velocity is lowest. As we have mentioned, one technique is to accelerate $H_2^+$ instead of protons [48]. The $H_2^+$ ion, produced by stripping one electron from a hydrogen molecule in the ion source, is stable and much more tightly bound than the $H^-$ ion (-2.6 eV and -0.7 eV, respectively). This renders the ion less susceptible to Lorentz stripping, but the vacuum in the cyclotron must still be quite high to avoid loss of beam from collisions of the ions with residual gas molecules that will dissociate the ion [49]. The advantage of $H_2^+$ is that one accelerates two protons, but experiences the space charge effect of only one charge.

The $H_2^+$ ion is produced in a similar ion source that one uses for generating proton ($H^+$) beams and is considerably simpler than the $H^-$ source. One creates a plasma with hydrogen gas, and extracts ions from an aperture at the end of the plasma chamber. The ionic constituents in the plasma will be $H^+$, $H_2^+$ and $H_3^+$, the ratio of these ions is finely controlled by the nature of the plasma: how it is generated, how electrons are injected into the plasma, and the total power level of the discharge. These parameters can be adjusted to maximize any one of the three. Our group at MIT has developed the MIST-1 source [50, 51], a multi-cusp, filament-driven source modeled after Ref. [52], with an $H_2^+$ fraction of over 80%. This source is, at present, operating on a test stand at MIT (Fig. 5), at current levels adequate to drive our proposed cyclotron. It is being upgraded to the MIST-2 configuration to increase its beam current -- by improving source cooling and adding plasma diagnostics to aid in optimizing performance -- and to provide a margin of safety in operation.

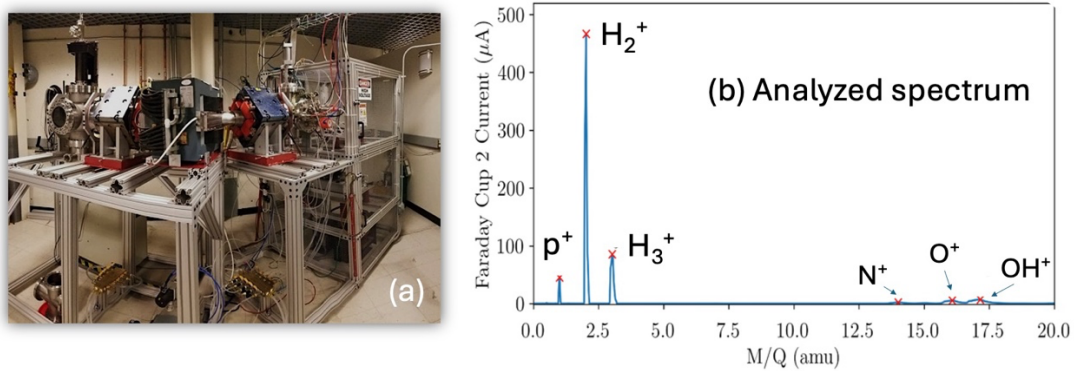

**Fig. 5** MIST-1 ion source. (a) photo of test stand: source is in high-voltage cage at far right, analysis magnet is at center, and instrumentation - emittance measurement and Faraday cup - on left. (b), spectrum showing high fraction of $H_2^+$ (5b from [50])

## Innovation 2: identify a more efficient bunching system

A second important point of development has been that of a more effective system of beam bunching. A widely used structure for low energy beam handling is the RFQ (Radio Frequency Quadrupole). Traditionally, the RFQ consists of four rods arranged as a quadrupole with a beam passing through the center along the axis of the rods. Radiofrequency is applied to the rods with opposite rods driven together so while one plane focuses the other defocuses, and when the phase is advanced by 180° the focusing planes reverse. This device has been used as an ion-beam mass analyzer for many years [53]. With smooth rods all the fields are transverse, so the longitudinal structure of the beam is not affected. But in 1970 it was realized that modulating the diameter of the rods by adding a (close-to) a periodic shape along their length could add longitudinal electric fields that could affect the bunching and acceleration of the beam [54]. In the late 1970's, a "Proof-of-Principle" machine was built at Los Alamos [55] where the continuous beam from an ion source was very efficiently bunched and at the same time accelerated by a modest amount. This innovation revolutionized the field of high-energy particle accelerators, replacing the large high-voltage terminals needed to give the beam sufficient energy to enter the first linear accelerator stage with a compact structure with a length on the order of meters and a diameter of around 50 cm. This structure also easily handles the space-charge forces in the low-energy beams: peak currents of the order of 100 mA are routine. The resonant electric field producing the very high gradients needed to manipulate the beam involved the excitation of transverse modes in the copper cavities, requiring frequencies of the order of 400 MHz.

The idea of using an RFQ as a buncher for a cyclotron was first proposed in 1982 [56] but was never implemented, largely because the buncher must operate at the same frequency as the cyclotron RF system, which is typically 30 to 40 MHz, i.e., a factor of ten lower than the typical RFQ structure frequencies. However, over the years different geometries of RFQ's have been designed, including the so-called split-coaxial structure [57] where the pairs of opposite rods are attached to the opposite end plates of the RFQ (i.e. the top and bottom rods are attached to the plate at the entrance to the RFQ, and the two side rods are attached to the exit plate), allowing the cavity to be excited in a longitudinal mode, which, because of the different sizes (length of the cylindrical cavity versus its diameter), can indeed be made to resonate in the correct frequency range.

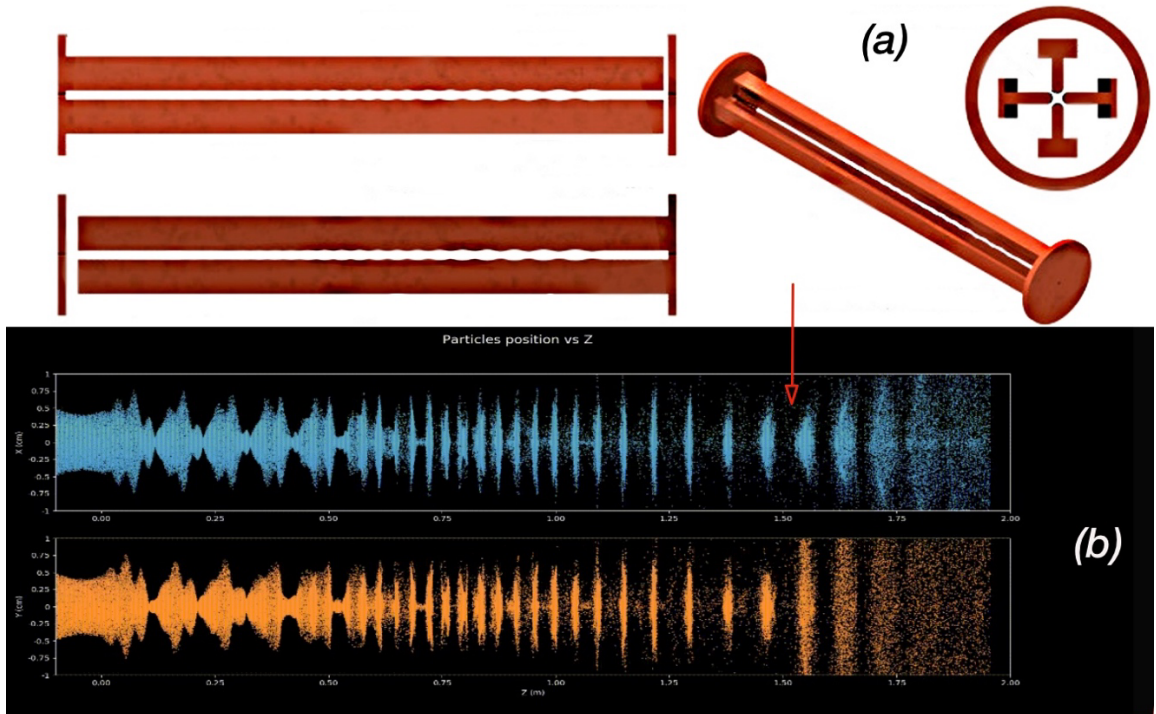

**Fig. 6** (a) Schematic of a "split coaxial RFQ" structure. The opposite pairs of rods (also called vanes) are attached to opposite end flanges. Note the modulation in the vanes that gently bunches and accelerates the CW beam entering from the left. (b) Particle-in-cell simulation showing how the beam is bunched, which is highest at the exit of the RFQ, denoted by the red arrow. Because of energy spread introduced the beam starts debunching right after the exit. The RFQ is placed as close to the inflection point as possible (see Fig. 10), to minimize the distance to the first accelerating cavity

We have designed an RFQ that is currently being constructed [58, 59], which achieves a bunching efficiency of over 90%. It is 1.49 meters long, 28 cm in diameter, and operates at our cyclotron frequency of 32.8 MHz. This bunching efficiency means that (allowing for collimation to clean up beam halo) about 60% of the injected beam will be captured in the central region. One should note, though, that the highest bunching factor is at or very close to the end of the RFQ, and because the structure adds some energy spread to the beam, it immediately begins to debunch after the end of the structure. It is necessary then to place the end of the RFQ as close as possible to the first accelerating gap in the cyclotron (denoted by the red arrow in Fig. 6). Fig. 10 shows the "front end" components, ion source and RFQ, axially mounted onto the cyclotron, with the RFQ inserted as close as possible to the plane of the cyclotron.

## Innovation 3: make use of vortex motion stabilization explicitly in the design

Particles presented to the accelerating gaps in the cyclotron RF system at the correct time (RF phase), will receive an accelerating kick that will direct them to the right location and time at the subsequent accelerating gaps. This is referred to as the "stable phase" which is approximately 30 degrees wide in a well-designed cyclotron. Particles in this region form into a bunch, that continues through the full acceleration cycle to the outer radius of the cyclotron. For the highest currents too, it is important that the cyclotron be "isochronous," that is the time it takes a bunch to make one full revolution around the cyclotron is the same regardless of the bunch radius or energy. This condition is automatically met in a flat magnet for non-relativistic particles, but for focusing, as well as relativistic considerations, the magnet is often broken up into sectors, referred to as "hills" for the narrow gaps (a few cm) with the highest magnetic fields, and "valleys" where the poles are far apart (many 10's of cm) which also allows for insertion of the RF "dees" with accelerating gaps at either edge. Isochronism is adjusted by changing the width of the hill sections as a function of radius. The important point is that the integral of the field around a full turn in the bunch's path meet the isochronicity condition. In an isochronous cyclotron, bunches can be found at all radii, again, providing a continuous stream of bunches to the extraction channel.

We have explained above that the stripping foil used in present-day isotope cyclotrons is a limiting factor in producing higher intensity beams [12]. Without foils, one can still use the original method for extracting beam from a cyclotron: placing a very thin metal sheet called a septum into the beam at the outer radius, with a plate on the outer side and connecting a high voltage supply between these two plates, generating an electric field that directs the beam away from the cyclotron. The kick is fairly gentle but is enough to create more separation between the last two turns to place magnetic components further along the extraction orbit that guide the beam out.

For this to work, one must have very clean turns in the cyclotron, with good separation between the last few turns, and the smallest possible bunches so the space between turns contains very few, if any, particles. As the septum sits at the midpoint between the last two turns, beam particles in this space have a high chance of striking the septum: causing - for our high currents - thermal damage to the septum, and activation of the cyclotron. Turn separation is obtained by using the highest possible RF voltage in the outer accelerating cavities, which is limited by voltage holding and breakdown[1]. Once the highest possible voltage is applied, the bunch size must be kept as low as possible by transverse focusing that is provided by the magnet pole shape and the dynamics of the particles through the accelerating gaps. This is the place where space charge becomes important. As the charge in each bunch is increased, the repulsive forces of the ions fight against the focusing forces, and one's intuition is that this would increase the equilibrium bunch size. However, a serendipitous discovery at the "Injector II" high current 72 MeV cyclotron at the Paul Scherrer Institute (PSI) [60] proved that the effect of space charge on these bunches was substantially more complex, and in fact led to stable bunches that did not grow in size. While the effect of vortex motion has first been observed many years ago, to our knowledge, PSI Injector II is the only cyclotron in existence that benefits from it. Our machine will be the first that is purpose-built to benefit from this effect. Recent theoretical treatments of vortex motion can be found in Refs. [61, 62, 63]. A simplified view for a uniformly charged, elliptical beam in a uniform magnetic field is shown in Fig. 7. Here the addition of velocity components tangential to the beam edge due to the interplay of self-forces (space charge) and the cyclotrons magnetic field can be seen. Careful simulations using the particle-in-cell OPAL code [64] show that this provides stability for the bunch which is established in the very first few turns (see Fig. 8). While halo is generated in this process, careful placement of a set of collimators can intercept these stray particles and provide a clean beam that is transported to the extraction radius with essentially no growth in size. The OPAL code was developed at PSI and has been used very successfully to simulate vortex motion in Injector II. To further build confidence in our design, we have performed robustness studies and uncertainty quantification with OPAL using machine learning techniques [65].

---

[1] Turn separation is also increased by structure resonances

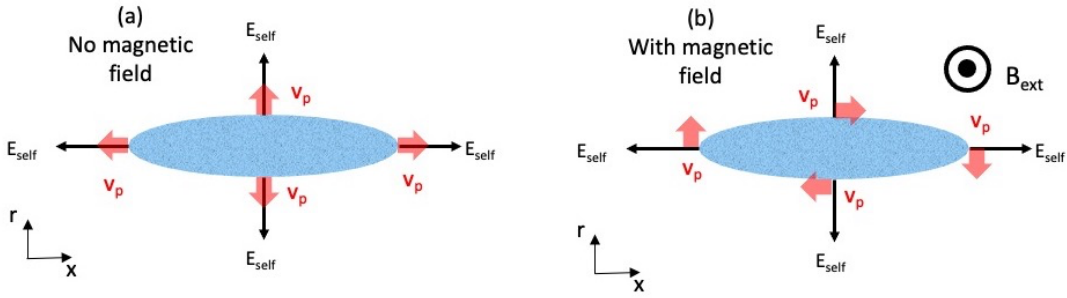

**Fig. 7** (a) Space-charge effect on a bunch as seen from above with no magnetic field, the bunch grows in size. (b) In the presence of the cyclotron magnetic field the expanding motion is bent into a vortex that remains stable to the outer radius of the cyclotron

Note that all the clean-up collimators are placed before the beam energy has reached 1.5 MeV/amu and nuclear effects will be negligible for an $H_2^+$ beam. Power deposited by the beam will be transferred away with DI (de-ionized) cooling water. Each collimator may be considered one half of a (radial) slit and will be adjustable a few mm radially.

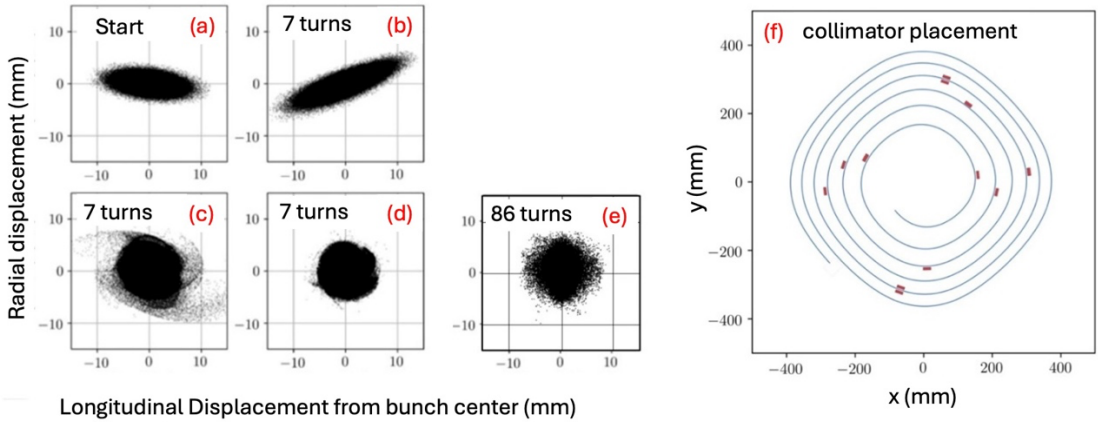

**Fig. 8** (a)-(e) OPAL simulations of a single bunch, each illustration viewed from above the cyclotron plane; (a) start, just after spiral inflector, (b)-(d) after 7 turns: (b) with no space charge, bunch evolves according to normal beam dynamics in the cyclotron, (c) after turning on the space-charge force, the effects of vortex motion are clearly apparent, approximately 40% of the beam particles are driven into the halo surrounding the main bunch, (d) collimators, shown in (f) remove this halo. (e) Follows the OPAL simulation to the extraction radius showing no growth in the beam size. (f) Shows the placement of the scraping collimators.

Further OPAL simulations indicate that the beam bunches are well separated at the location of the extraction septum. Fig. 9 shows the predicted distribution of ions at the outer edge of the cyclotron. The extraction septum is shown in red. Note the y axis is logarithmic, so the amount of beam that is predicted to hit the septum is very low. For the full 5 mA of $H_2^+$, it is predicted that the septum will absorb at most 100 watts of power. Experience at PSI has been that beam losses of less than 200 watts provide a low enough activation environment to allow for hands-on maintenance of their cyclotrons.

A small cyclotron (shown in Fig. 10) called the HCHC-1.5 (High-Current $H_2^+$ Cyclotron) is currently being constructed that will include both the ion source and RFQ and will inject and accelerate beam through 5 full turns to an energy of 1.5 MeV/amu, low enough to avoid neutron production and activation, but high enough to demonstrate the establishment of stable bunches through the vortex motion effect.

### Tuning strategy for the IsoDAR machine

To mitigate thermal and activation effects on the collimators, the total beam current will be severely reduced during the initial injection and tuning studies. This is done by use of a fast chopper in the injection line before the RFQ, that will ultimately reduce the number of bunches residing in the cyclotron without affecting the density of particles within each bunch. It is important to preserve this bunch density to establish the vortex effect, while reducing the number of bunches might only be important if there are substantial inter-bunch interactions. This can be tested too by keeping every 4[th] bunch[2], so that all bunches along a radial line are resident in the machine, and still reducing the total beam current by a factor of 4. As the collimator placement is refined and the tuning parameters established, the number of bunches can be increased gradually to ensure proper behavior of the overall central-region system.

---

[2] The cyclotron operates in the "4th harmonic", i.e. the RF frequency is 4 times the particle revolution frequency, so there are four stable bunches in the same orbit around the circumference, 90° apart.

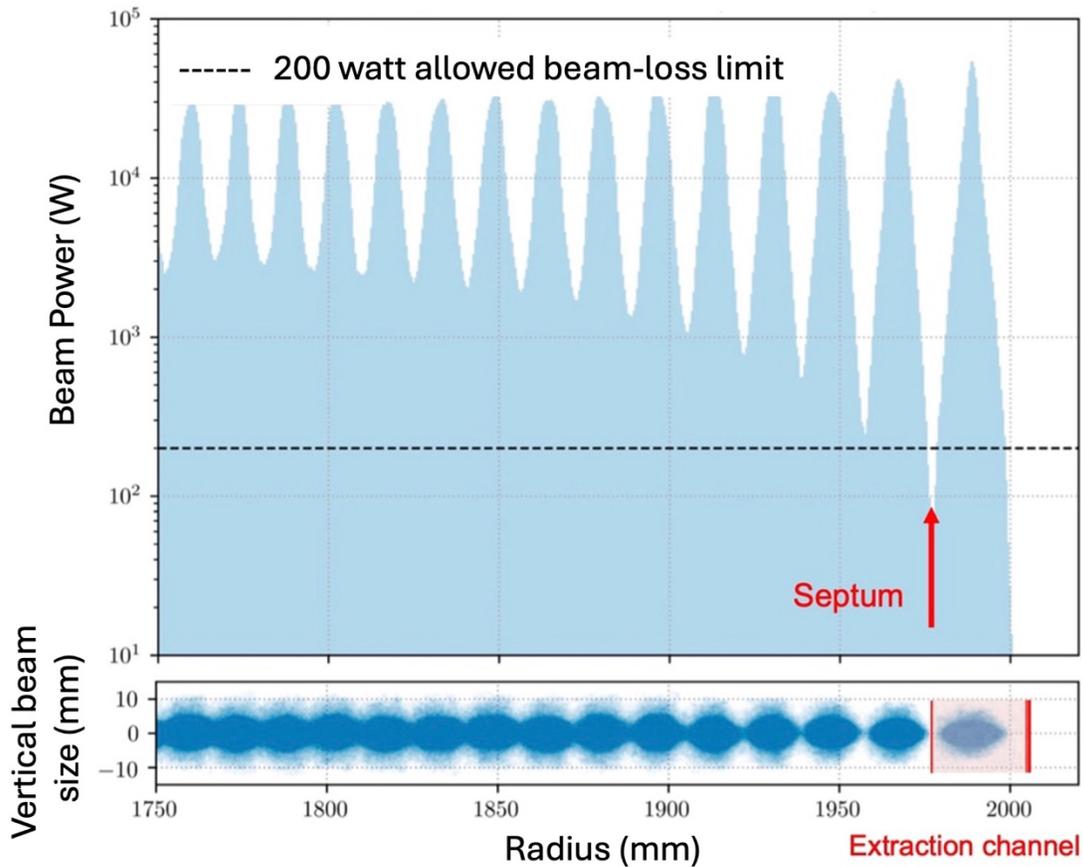

**Fig. 9** OPAL simulation of beam sizes for outer regions of the cyclotron. Excellent turn separation is predicted, with very low losses on the extraction septum. (Adapted from [65])

## Extending Applications: Flexibility and Constraints

As has been shown, the innovations we have developed really occur in the injection process to achieve efficient bunching thus reducing beam loss in the central region, and in the first few turns to establish the vortex motion that stabilizes the bunches. Beyond these, one has the option of placing the outside edge of the magnet, determining the cyclotron size, at whatever value and beam energy is desired for a particular application. Fig. 10 shows, for instance, two examples of cyclotrons of greatly different sizes: the IsoDAR cyclotron (HCHC-60) with an outer diameter of 6 meters, and the much smaller HCHC-1.5 (also called the "Demonstrator") about 1 meter in diameter, with a beam accelerated to 1.5 MeV/amu. Both have the same injection line and transport to the median plane of the cyclotron. We can see this flexibility having potential value in adapting these concepts to tailoring isotope-producing cyclotrons for specific needs: all with very high currents, and beam-energy or ion species chosen for each application.

*Comments*

Note that these cyclotrons, specified to accelerate $H_2^+$, will be larger than corresponding $H^-$ cyclotrons producing the same energy of proton beams, because of the higher magnetic rigidity of the $H_2^+$ ion with its charge-to-mass ratio of 1 to 2.

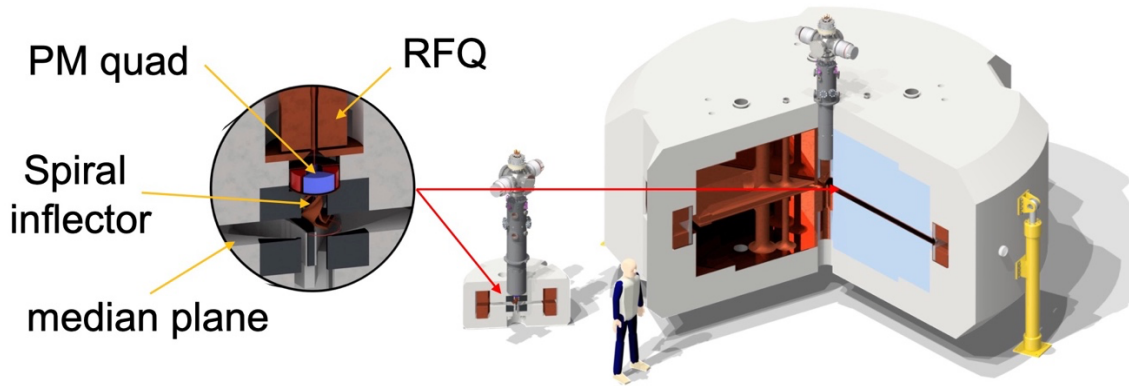

**Fig**. **10** Two members of the new family of cyclotrons. The 60 MeV/amu IsoDAR cyclotron (HCHC-60), and the 1.5 MeV/amu HCHC-1.5. Both use identical injection lines. The permanent magnet quad provides transverse focusing between the RFQ and the spiral inflector.

A side note on units and ways of expressing beam energy might be in order here. An accelerator scientist is interested in the ion velocity, namely the time it takes to travel from one RF cavity to the next. This person will usually express the energy of a particle as "MeV/amu," or energy per atomic mass unit. Regardless of the total mass of each ion, ions with the same MeV/amu will all have the same velocity. A nuclear physicist on the other hand is interested in the total energy of a projectile that is reacting with a target nucleus, so will count both protons and neutrons. For instance, an $H_2^+$ ion in the IsoDAR cyclotron will be extracted at 60 MeV/amu, each proton has a kinetic energy of 60 MeV. If this ion is dissociated, by passing it through a stripper foil, for instance, the resulting beam will have twice the total electrical current. Thus 5 milliamps of 60 MeV/amu $H_2^+$ after the stripper will be 10 milliamps of 60 MeV protons. On the other hand, a deuteron accelerated in the IsoDAR cyclotron will have an extracted energy of 60 MeV/amu, but this means that both the proton and the neutron have 60 MeV, so the total energy of the ion is 120 MeV.

If the powerful $H_2^+$ beam is stripped after extraction with a septum, then the foil is not in a magnetic field, so the single convoy electron is not bent back into the foil. In this case, the foil lifetime is not related to heating, but only to dislocation damage due to displacements of the carbon atoms from close encounters between the high-energy protons and the carbon nuclei. We performed a test at PSI, placing a foil in the 72 MeV transport line between the Injector II and the main ring cyclotrons, running a 1.7 mA beam through the foil for over 60 hours [66]. At the end of the run the foil did show signs of damage -- stretching and distortions -- but had no perforations.

## Conclusions

The innovations we have developed over the past 10 years, summarized in a recent bibliography of our publications [67], have the potential to carry the compact cyclotron to a new plateau of beam-current performance, which can have tremendous impacts, not only on fields such as neutrino science, but also potentially in isotope production. For example, for production of long-lived generator parents, and in addition, cyclotrons accelerating deuterons to energies around 40 MeV (20 MeV/amu) could be the workhorses meeting the world's demand for $^{225}$Ac. We are actively continuing R&D efforts and the building of cyclotron prototypes to bring these concepts to reality. We encourage the isotope community to consider the value of these novel new cyclotrons and address R&D issues that will allow their use.

## Acknowledgements

We would like to acknowledge Lee Bernstein for discussions relating to fast neutron production and their potential for effective utilization of our high-intensity beams. We also wish to acknowledge the IBA Accelerator R&D team and the members of the cyclotron and simulation groups within the IsoDAR Collaboration for their enthusiastic support of this work and extremely valuable assistance.

## Declarations

**Funding:** This work was supported by NSF grants PHY-1707969, PHY-1912764 and PHY-1626069, DOE grant DE-SC0024138, and the Heising–Simons Foundation.

**Data Availability Statement:** Supporting data are available from the authors upon reasonable request.

**Conflicts of Interest:** The authors declare no conflict of interest.